\documentstyle[emulateapj]{article}

\submitted{Received 1998 October 2; accepted 1998 November 21}

\received{2 October 1998}
\accepted{21 November 1998}
\lefthead{Zubko, Smith, \& Witt}
\righthead{Silicon nanoparticles and interstellar extinction}

\begin{document}

\title{Silicon nanoparticles and interstellar extinction}

\author{Victor G. Zubko\altaffilmark{1}}
\affil{Department of Physics, Technion -- Israel Institute
             of Technology, Haifa 32000, Israel; \\
             zubko@phquasar.technion.ac.il}

\and

\author{Tracy L. Smith and Adolf N. Witt}
\affil{Ritter Astrophysical Research Center, University of Toledo,
             Toledo, OH 43606; \\
             tsmith@astro1.panet.utoledo.edu, awitt@astro1.panet.utoledo.edu}

\altaffiltext{1}{On leave from the Main Astronomical Observatory, NAS,
                 Kiev, Ukraine}

\begin{abstract}
To examine a recently proposed hypothesis that silicon nanoparticles
are the source of extended red emission (ERE) in the interstellar medium,
we performed a detailed modeling of the mean Galactic extinction in
the presence of silicon nanoparticles. For this goal we used the appropriate
optical constants of nanosized Si, essentially different from those of bulk
Si due to quantum confinement. It was found that a dust mixture of silicon
nanoparticles, bare graphite grains, silicate core-organic refractory mantle
grains and three-layer silicate-water ice-organic refractory grains works
well in explaining the extinction and, in addition, results in
the acceptable fractions of UV/visible photons absorbed by silicon
nanoparticles: 0.071--0.081. Since these fractions barely agree with
the fraction of UV/visible photons needed to excite the observed ERE,
we conclude that the intrinsic photon conversion efficiency of
the photoluminescence by silicon nanoparticles must be near 100\%,
if they are the source of the ERE.
\end{abstract}


\section{Introduction}

Recently, extended red emission (ERE) has been detected in
the diffuse galactic background and in high-galactic latitude cirrus clouds
with surprisingly high intensities (Gordon, Witt, \& Friedmann 1998;
\cite{sg}). This emission, appearing in the 500--800 nm spectral range as
an unstructured broad feature, is attributed to photoluminescence
by some component of interstellar grains. The correlation between the ERE
R-band intensity and \ion{H}{1}-column density of (1.43$\pm$0.31)\,10$^{-29}$
ergs s$^{-1}$ {\AA}$^{-1}$ sr$^{-1}$ H atom$^{-1}$  (\cite{gwf})
implies that, at a minimum, (10$\pm$3)\% of all UV/visible
photons in the diffuse interstellar radiation field are absorbed by
the ERE producing grains, if each absorption results in the emission of
an ERE photon. This requires that the ERE-grains make a significant
contribution to the interstellar extinction in the UV/visible range,
which further implies that they must consist of cosmically abundant
elemental constituents in a form that permits extremely high efficiency
photoluminescence.

After a review of possible ERE-grain candidates, Witt, Gordon,
\& Furton (1998) and Ledoux et al. (1998) independently proposed that silicon
nanoparticles (SNPs) might be the source of ERE in the interstellar medium
in as far as they produce photoluminescence spectra closely matching those
observed in ERE sources and match the constraints regarding the required
quantum efficiency and elemental abundance. It was discovered experimentally
that the photoluminescence emission originates from nanocrystalline Si regions
in porous Si material (\cite{canham}; \cite{cole}) and from clusters of
nanosized grains (\cite{takagi}; \cite{littau}; \cite{tamir}). The estimate
of the amount of silicon needed in the form of SNPs, however, was based upon
the average UV absorption cross section of crystalline silicon in bulk form.
This must be considered as only a rough approximation, because the remarkable
optical properties of SNPs, clusters involving a few 10$^2$ silicon atoms
each with a silicon suboxide (SiO$_x$) surface covering, arise precisely
because quantum confinement alters the band structure of the cluster material
relative to that in the bulk material. Since quantum confinement is a function
of cluster size, so are the resulting photoluminescence characteristics,
and so are the dielectric functions of the cluster material.

The current investigation was undertaken to answer several related
questions: What are the UV/optical extinction characteristics of 
a separate component of SNPs? What fraction of the silicon
believed to be depleted into solid grains is needed in the form of SNPs 
to explain the observed ERE intensities? What resulting
characteristics of the interstellar extinction curve can be related to
absorption by SNPs and can thus be correlated with varying
ERE intensities? To find answers on these questions, we made use of
a powerful computational approach based on the method of regularization
(\cite{zubko}), which was proven to be efficient in modeling
of interstellar dust (Zubko, Kre{\l}owski, \& Wegner~1996,~1998).

\section{Optical properties of Si nanoparticles}
     \label{sec:opt_conts}

The first step in our study was to choose the optical constants
of SNPs. Witt et al. (1998) proposed that the interstellar SNPs should
have diameters of 1.5--5 nm because only grains in this size range exhibit efficient photoluminescence. Evidently, the dielectric properties of such
SNPs should be different from those of a bulk material because of
both the quantum size confinement effect and the increased role of
surface states. These expectations are fully confirmed by available
experimental data (\cite{koshida}; \cite{theiss}).
In particular, Thei{\ss} (1997) reported the dielectric function
of nanosized silicon from porous samples with a range
of porosities 60--80\%, corresponding to the respective range of
crystallite sizes of 5--2 nm. These results show that
the optical properties of silicon nanostructures have a strong
size dependence. However, an accurate derivation of the latter is not
a trivial problem, because in each silicon sample there is some
size distribution of crystallites (\cite{theiss}).
Note that Thei{\ss}' (1997) dielectric function covers the wavelengths
0.2--1 $\mu$m only. However, as we will see below, we especially need the data
down to the Lyman limit (912~{\AA}). We therefore decided to use in our
modeling the optical constants of nanosized silicon by Koshida et al. (1993)
spanning the range 0.055--0.85 $\mu$m, which is sufficient for our problem.
These constants correspond to the characteristic diameter of a SNP of around 3 nm. The optical constants of nanosized silicon by Koshida et al. (1993) and of
bulk silicon taken from the compilation by Palik (1991) are displayed
in Fig.~{\ref{fig:oc}}. There is a considerable difference between the two
sets which may also be seen in the extinction efficiencies of the respective
spherical grains shown
in Fig.~{\ref{fig:qext}}. The first three cases correspond to the SNPs used
in our modeling (see Section 3 for more details). 
Curves 3 and 4 demonstrate the so-called {\em extrinsic size effect}
for small particles when the bulk optical constants are used.
The peak at 0.118 $\mu$m is due to the SiO$_2$ shell and is present also
in the spectra of bulk material, whereas the peaks at 0.153 $\mu$m and
0.138 $\mu$m (for 3 and 4, respectively) are surface plasmons
due to the silicon core, specific for small particles. The peak position
depends on the volume fraction of Si. In contrast to the latter cases,  
curve 1 shows the {\em intrinsic size effect} when
the optical constants themselves vary with grain size. Note that
Draine \& Lee (1984) already computed the size-dependent
dielectric function of small graphite grains. However, the size dependence
was taken into account in their calculation only for the free-carrier part
of the dielectric function, whereas the part due to the interband transitions
was set to be the same as for bulk graphite. In contrast to Draine \& Lee 
(1984), the optical constants of nanosized silicon used in the present study
explicitly reflect the drastic changes in the band structure and interband
transitions of the SNPs in comparison with those of bulk silicon.

\section{Interstellar extinction with Si nanoparticles}
       \label{sec:extinction}

The dust models investigated for the present study were expected to satisfy
three independent constraints simultaneously: 1) fit the mean Galactic
extinction curve as closely as possible over the entire near-IR to far-UV
range; 2) consume depleted heavy elements consistent with the recently
proposed reference abundances for the interstellar medium (about 2/3 of
solar values, see e.g., Snow \& Witt 1996); and 3) require the
ERE-producing SNPs to absorb at least 0.10$\pm$0.03 of the UV/visible
photons in the interstellar radiation field, as suggested by the observations
of Gordon et al. (1998). The problem was solved by using the approach
and computer program based on the method of regularization
(\cite{zubko}; \cite{zkw2}) which allows one to find the size distributions
of the dust components in the most general form. The size distributions may
be subject to the cosmic abundance and mass fraction constraints.
Following Zubko et al. (1998), we treated the mixtures of spherical bare,
multilayer and/or composite grains, whose extinction efficiencies were
calculated either with the standard Mie theory for bare and composite grains
(Bohren \& Huffman 1983) or with its extension to multilayer particles.
For composite grains the effective dielectric function was first calculated
using the effective medium approach by Stognienko, Henning, \& Ossenkopf (1995).

The fraction of UV/visible photons absorbed by SNPs, $\beta$, was calculated
with the formula
\begin{equation}
\beta =
 {  { \int_{0.0912 {\mu}m}^{0.55 {\mu}m} (1-a_{\lambda}^{\rm{SNP}}) 
           \,{\tau}_{\lambda}^{\rm{SNP}} I_{\lambda}^{\rm{ISRF}}
           \,\rm{d}{\lambda}
    }
      \over 
    { \int_{0.0912 {\mu}m}^{0.55 {\mu}m} (1-a_{\lambda}) 
           \,{\tau}_{\lambda} I_{\lambda}^{\rm{ISRF}}
           \,\rm{d}{\lambda}
    }
 }
       \label{eq:beta}
\end{equation}
where $a_{\lambda}$ ($a_{\lambda}^{\rm{SNP}}$) is the albedo of the
entire dust mixture (SNPs), ${\tau}_{\lambda}$ (${\tau}_{\lambda}^{\rm{SNP}}$)
is the extinction of the entire dust mixture (SNPs), $I_{\lambda}^{\rm{ISRF}}$
is the intensity of the interstellar radiation field at the Earth
in units of photons cm$^{-2}$ s$^{-1}$ {\AA}$^{-1}$. The $a_{\lambda}$
was taken from Gordon et al. (1998), ${\tau}_{\lambda}$,
the mean Galactic extinction ($R_V$=3.1) from Cardelli, Clayton, \& Mathis
(1989) and $I_{\lambda}^{\rm{ISRF}}$ from Mathis, Mezger, \& Panagia (1983).
The values of $a_{\lambda}^{\rm{SNP}}$ and ${\tau}_{\lambda}^{\rm{SNP}}$
were taken from the models.

We based our calculations on the G, M and O types of models introduced
by Zubko et al. (1998). A G model contains the constituents usual for
Greenberg's models: the bare graphite grains and the silicate core--organic
refractory grains (e.g. Li \& Greenberg 1997). An M model is a mixture of
the bare graphite and silicate grains together with the porous composite grains,
resembling recent Mathis' (1996) models. The constituents of an O model are
the bare graphite and silicate grains together with the multilayer grains
consisting of a silicate core coated by refractory organic and water-ice shells.

Initially, we tried to model the extinction using the above models to which
we added SNPs of 3.0 nm diameter covered by a SiO$_2$ mantle. We varied
the thickness of the mantle and found that the best results were achieved
when the mantle is quite thin. However, all the models resulted in quite low
$\beta$, as a rule not exceeding 0.03--0.04, and, as a rule, consumed
the solar fractional amount of silicon, 35 ppm (atoms per 10$^6$ H atoms).
For the next step we introduced spherical porous clusters of SNPs instead of
isolated SNPs. This resulted in an increase of $\beta$, typically by
0.003--0.005 (the highest values correspond to 70\% porosity). Then we tried
combining the G, M, and O models and found that the only mixture
that results in acceptable $\beta$ ($>$ 0.07) is a combination
of G and O models or the GO model. It consists of four components:
bare graphite grains, silicate core-organic refractory mantle grains,
three-layer silicate-water ice-organic refractory grains and, finally,
the SNPs in the form of either silicon core-SiO$_2$ mantle grains or
70\%-porous aggregates of silicon grains containing a small amount
of SiO$_2$. The GO model requires a mass fraction of 7\% of silicon grains.
It results in $\beta$=0.071 and 0.081 for the models with isolated
and clustered SNPs, respectively. Note, however, that these models still
consume much carbon, 190 ppm, and a solar fractional amount of silicon.
The GO model seems more natural than G and O models separately,
because when the dust grains in the interstellar medium do evolve
through the formation and processing of mantles, we should
expect the contributions into extinction of various intermediate stages.

\placefigure{fig:models}

\placetable{tab:tab1}
\placetable{tab:tab2}

The main results of our calculations are presented in
Fig. \ref{fig:models} and Tables \ref{tab:tab1}--\ref{tab:tab2}.
First of all, we see that the models with bulk silicon exhibit rather
strong features in the UV (see Fig. \ref{fig:qext}) which so far have not
been observed. For this case we adopted the volume fractions:
$\delta_{\rm{Si}}$=0.297
and $\delta_{\rm{SiO_2}}$=0.703 corresponding to the initial assumption
of Witt el al. (1998) about an equal amount of silicon and oxygen
atoms in a SNPs. By contrast, the contribution of the SNPs to the
extinction is featureless when the optical constants of nanosized silicon
and a very thin SiO$_2$ mantle are used. As a result, such models fit
the mean Galactic extinction much better.
Practically all of the models require much carbon: 180--190 ppm,
except O models for which C/H=100 ppm, which is about optimal.
As found in previous models by Zubko et al. (1998), more oxygen is
needed for the O models. The amount of silicon needed requires solar
relative abundances.
A remarkable feature of the GO models in contrast to the other models is
the extremely small amount of magnesium and iron consumed,
about 17 ppm, which is quite close to the respective optimum values by
Snow \& Witt (1996). In addition to providing the highest $\beta$ values,
the GO models also show the closest fit to the observed mean Galactic 
extinction curve.

A result found to be generally true for all of our models is
that $\beta$ is approximately equal to the mass fraction incorporated in the 
SNPs (see Table \ref{tab:tab2}). This result is largely independent of whether
the SNPs are included as a single-size component of $a$=1.5 nm, an $a^{-3}$
distribution of porous SNP clusters in the range 1.5--50 nm, or, as pointed
out by the referee, if SNPs are included as the small-size tail of an
$a^{-3.5}$ distribution, as long as their mass remains constant.
Since the cosmic abundance of Si limits
this mass fraction severely and approximately 50\% of the available Si is
already incorporated in the SNPs in the GO models, one important conclusion
to draw is that if ERE is produced by SNPs, the intrinsic photon conversion
efficiency must be extremely high, essentially near 100\%. 

One reason why the $\beta$ values from Table \ref{tab:tab2} are as small
as they are is that the number of photons in the assumed radiation field
is highest in the 0.35--0.55 $\mu$m wavelength range, where the absorption
contribution of the SNPs is practically zero. One would obtain larger
values of $\beta$, if the interstellar radiation field contained increased
numbers of UV photons. If one can make a case for significant differences
in the spectral energy distribution of the interstellar radiation field
in different regions of the Galaxy, one should be able to predict
corresponding variations in the relative brightness of the ERE and
the visible diffuse galactic light (DGL). Such variations should be apparent
in results from ongoing surveys which map the DGL and the ERE over
a wide range of galactic longitudes (e.g., \cite{gwf}).
  
We conclude that in general the results of our models
are consistent with the hypothesis that SNPs are the source
of ERE in the interstellar medium. Their extinction contribution occurs
primarily at wavelengths shorter than 300 nm and is not expected to introduce
discernable structure into the interstellar extinction curve, provided one
takes into account the size dependence of the optical constants for
nanostructured silicon. The intrinsic photon conversion efficiency of such 
SNPs must be near 100\%, if cosmic abundance constraints are to be met. 
Spherical clusters of SNPs with cluster radii up to 50 nm provide a slightly
better match to the ERE requirement than do single SNPs with the same total
mass fraction. It would be of great interest to predict possible spectral
features due to SNPs in the infrared, as well as the contribution of SNPs
in the total thermal emission of interstellar dust. However, we cannot
do this because of the lack of optical constants of SNPs
in the infrared and because of lacking information about the size distribution
of SNPs.

\begin{acknowledgments}
We thank Prof. N. Koshida for providing us the electronic table
of the optical constants of nanosized silicon.
\end{acknowledgments}

\begin{deluxetable}{cclrrrrrrr}
%\tablewidth{0cm}
\tablefontsize{\small}
\tablecaption{The parameters of the GO models. \label{tab:tab1}}
\tablehead{
  \colhead{D\tablenotemark{a}} &
  \colhead{T\tablenotemark{b}} &
  \colhead{Components}         &
  \colhead{$f_{\rm{mass}}$\tablenotemark{c}} &
  \colhead{C\tablenotemark{d}}  &
  \colhead{Si\tablenotemark{d}} &
  \colhead{O\tablenotemark{d}}  &
  \colhead{Mg\tablenotemark{d}} &
  \colhead{Fe\tablenotemark{d}}
}
\startdata
n & i &               & 100.0 & 190 & 35 & 180 & 17 & 17\nl\nl

  &   & Graphite      &  22.1 & 143 &  0 &   0 &  0 &  0 \nl
  &   & Sil-ORR       &  40.0 &  33 & 12 &  83 & 12 & 12 \nl
  &   & Sil-Ice-ORR   &  30.9 &  14 &  5 &  97 &  5 &  5 \nl
  &   & Si-SiO$_2$    &   7.0 &   0 & 18 &   0 &  0 &  0 \nl\nl
n & c &               & 100.0 & 190 & 35 & 180 & 17 & 17 \nl\nl
  &   & Graphite      &  22.1 & 143 &  0 &   0 &  0 &  0 \nl
  &   & Sil-ORR       &  40.0 &  33 & 12 &  83 & 12 & 12 \nl
  &   & Sil-Ice-ORR   &  30.9 &  14 &  5 &  96 &  5 &  5 \nl
  &   & Si-SiO$_2$-voids & 7.0 &  0 & 18 &   1 &  0 &  0 \nl\nl
b & i &               & 100.0 & 190 & 29 & 180 & 17 & 17 \nl\nl
  &   & Graphite      & 23.5  & 145 &  0 &   0 &  0 &  0 \nl
  &   & Sil-ORR       & 40.0  &  32 & 12 &  80 & 12 & 12 \nl
  &   & Sil-Ice-ORR   & 29.5  &  13 &  5 &  88 &  5 &  5 \nl
  &   & Si-SiO$_2$    &  7.0  &   0 & 12 &  12 &  0 &  0 \nl
\enddata

\tablenotetext{a}{The optical constants of nanosized (n) or bulk (b) silicon.}
\tablenotetext{b}{The type of a SNP: isolated (i) or clustered (c).}
\tablenotetext{c}{The mass fractions of the dust components ($f_{\rm{mass}}$).}
\tablenotetext{d}{The amounts of chemical elements in dust in ppm:
   atoms per 10$^6$ H atoms.}
\end{deluxetable}

\begin{deluxetable}{cccccc}
\tablefontsize{\small}
%\tablewidth{0pt}
\tablecaption{The parameters of the models.  \label{tab:tab2}}
\tablehead{
  \colhead{Model} &
  \colhead{D} &
  \colhead{T} &
  \colhead{$\bar \mu$} &
  \colhead{$f_{\rm{SNP}}$} &
  \colhead{$\beta$}
}
\startdata
GO    & n & i & 0.284 & 0.070 &  0.071 \nl
GO    & n & c & 0.289 & 0.070 &  0.081 \nl
GO    & b & i & 0.406 & 0.070 &  0.054 \nl\nl

G     & n & i & 0.297 & 0.030 &  0.033 \nl
G     & n & c & 0.273 & 0.030 &  0.038 \nl
G     & b & i & 0.315 & 0.030 &  0.026 \nl\nl

M     & n & i & 0.352 & 0.020 &  0.021 \nl
M     & n & c & 0.353 & 0.020 &  0.024 \nl
M     & b & i & 0.357 & 0.020 &  0.018 \nl\nl

O     & n & i & 0.373 & 0.033 &  0.037 \nl
O     & n & c & 0.418 & 0.033 &  0.042 \nl
O     & b & i & 0.427 & 0.033 &  0.029 \nl
\enddata

\tablenotetext{a}{The mass fractions ($f_{\rm{SNP}}$) and the fractions of
   UV/visible photons absorbed by SNPs ($\beta$).}
\tablenotetext{b}{The quality of the fit is characterized by the normalized measure
   of incompatibility $\bar \mu$ (Zubko et al. 1996).}
\end{deluxetable}

\plotone{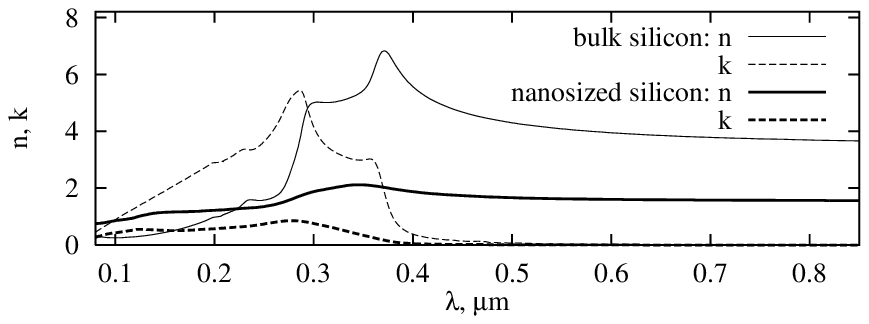}
\figcaption[nk.eps]{The optical constants of bulk and nanosized silicon.
         \label{fig:oc}
}

\plotone{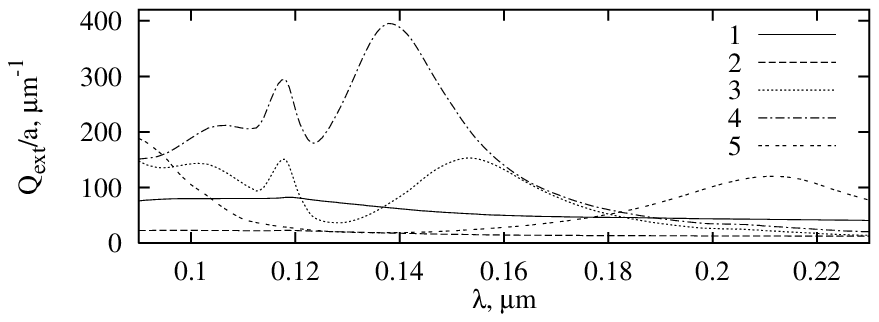}
\figcaption[qext.eps]{The extinction efficiency per grain radius,
   $Q_{\rm{ext}}/a$, of the SNPs and graphite grains.
   The optical constants of both nanosized (NS) and bulk (BS) silicon
   were used. Shown are the cases: 1: NS (0.99) core-SiO$_2$ (0.01) mantle,
   2: NS(0.29), SiO$_2$ (0.01), voids (0.7) in a composite grain,
   3: BS (0.297) core-SiO$_2$ (0.703) mantle,
   4: BS (0.700) core-SiO$_2$ (0.300) mantle, and
   5: graphite grains.
   The respective volume fractions are noted in brackets.
       \label{fig:qext}
}

\clearpage

{\plottwo{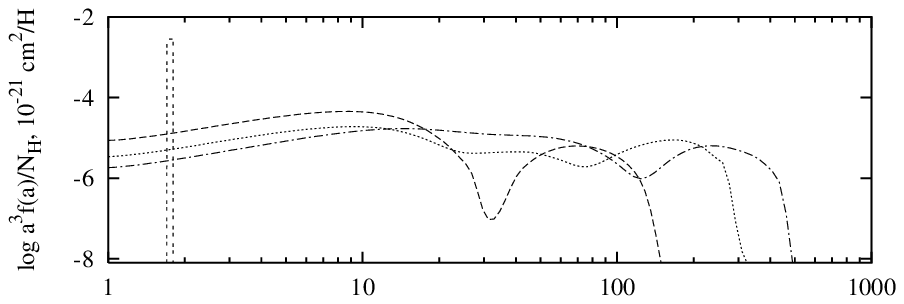}{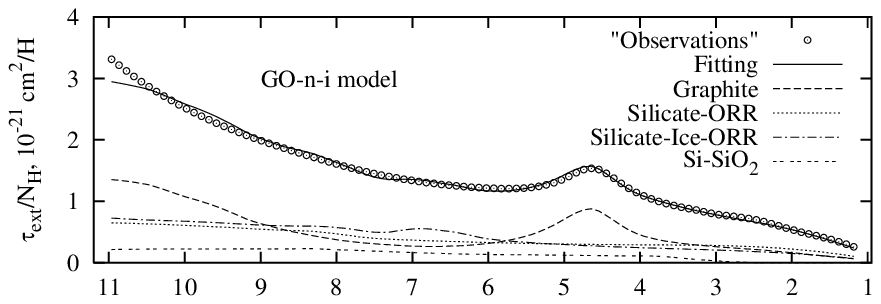}}
{\plottwo{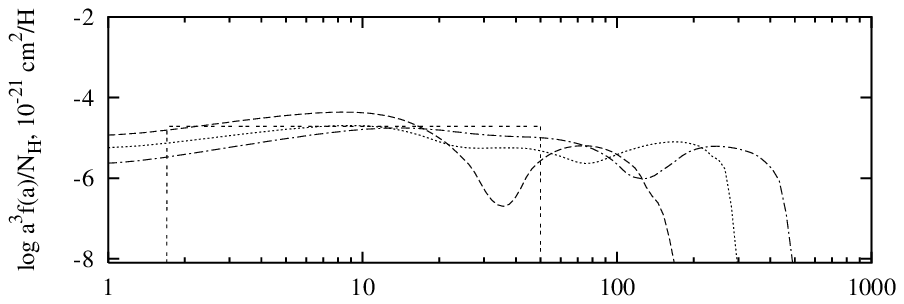}{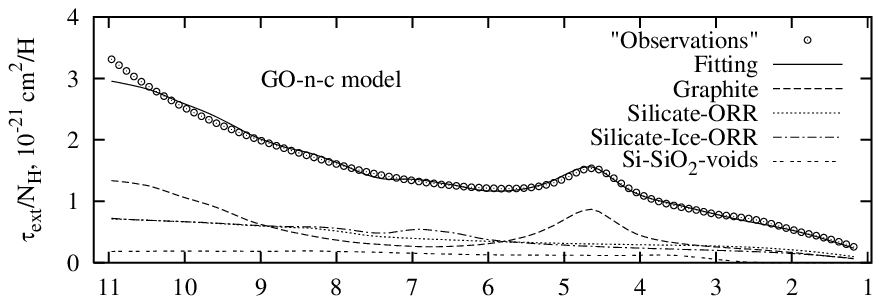}}
{\plottwo{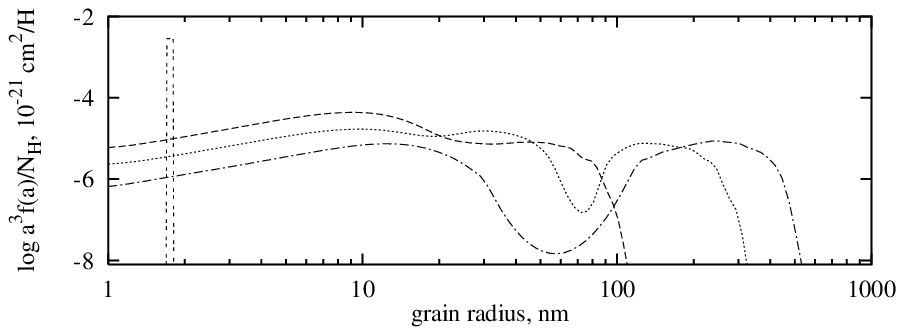}{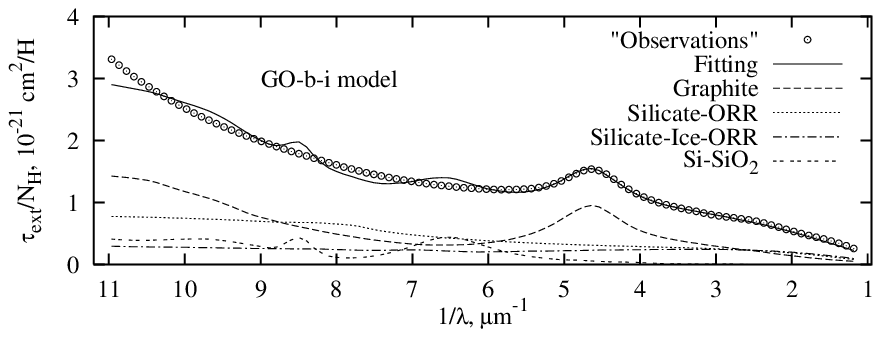}}
\figcaption[df3-1.eps ec3-1.eps df3-2.eps ec3-2.eps df3-3.eps ec3-3.eps]{
   Size distributions of dust grains (left panels) and
   respective extinction curves (right panels), fitting  the Galactic
   mean extinction curve ($R_V$=3.1). Shown are the GO-type dust models.
   See the text and Tables \ref{tab:tab1}--\ref{tab:tab2} for more details.
       \label{fig:models}
}
\clearpage

\end{document}